\documentclass[preprint,nopreprintline,12pt]{elsarticle}
\usepackage{amsmath, float}
\usepackage[caption=false]{subfig}

\begin{document}
\begin{frontmatter}

\title{Statistics of grain growth: experiment versus the Phase-Field-Crystal and Mullins models}

\author[1]{Gabriel Martine La Boissoni\`ere\corref{cor1}}
\ead{gabriel.martine-laboissoniere@mail.mcgill.ca}

\author[1]{Rustum Choksi}

\author[2]{Katayun Barmak}

\author[3]{Selim Esedo\={g}lu}

\cortext[cor1]{Corresponding author.}
\address[1]{Department of Mathematics and Statistics, McGill University, 805 Rue Sherbrooke O, Montr\'eal, QC H3A 0B9, Canada}
\address[2]{Department of Applied Physics and Applied Mathematics, Columbia University, New York, NY 10027, USA}
\address[3]{Department of Mathematics, University of Michigan, 530 Church St. Ann Arbor, MI 48105}

\begin{abstract}
We present a detailed comparison of multiple statistical grain metrics for previously reported experimental thin film samples of aluminum with 2D simulations obtained from the Phase-Field-Crystal (PFC) model and a Mullins grain boundary motion model. For all these results, ``universality'' is observed with respect to the dynamics and initial conditions. This comparison reveals that PFC reproduces geometric metrics such as area and perimeter, but does not capture grain shape and topology as accurately. Similarly, the Mullins model captures the number of sides distribution quite well but not other metrics. Our collective comparison of such measurements underscores the critical importance of the use of multiple metrics for comparison of experiments with all present and future models of grain growth in polycrystalline materials.
\end{abstract}


\end{frontmatter}

\section{Introduction}
\begin{figure}[h]
	\centering\subfloat[]{\includegraphics[width=0.24\textwidth]{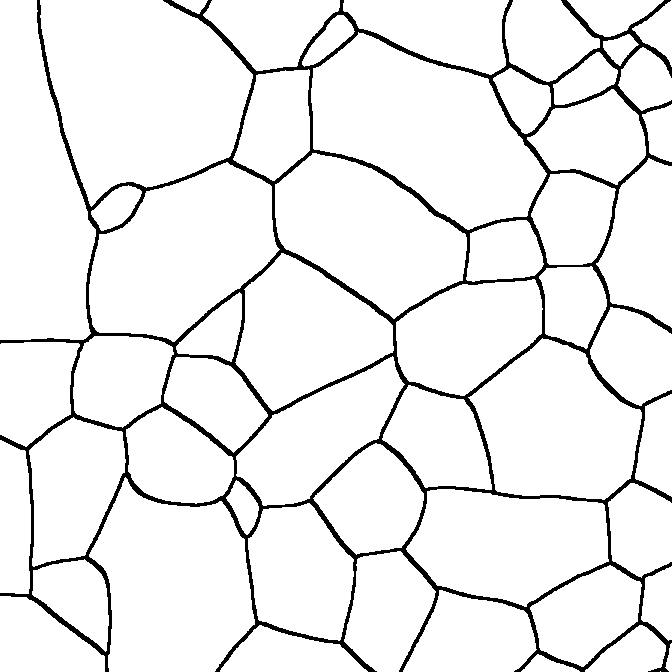}}
	\qquad \subfloat[]{\includegraphics[width=0.24\textwidth]{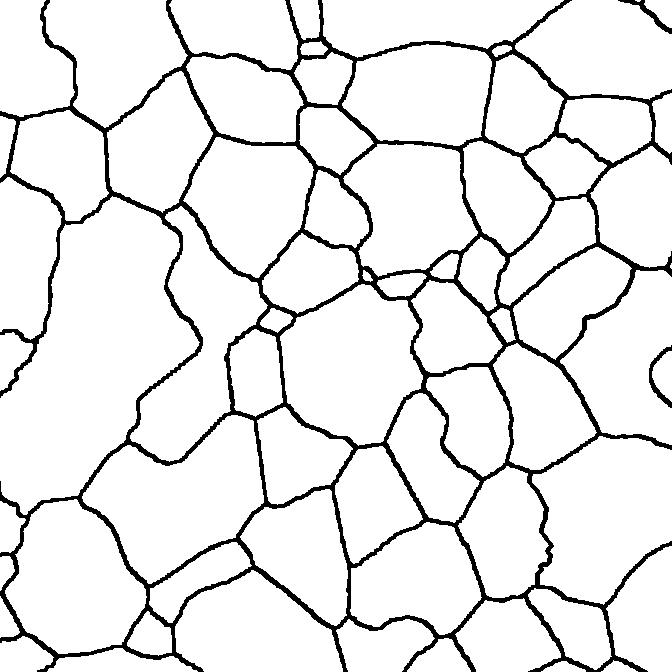}}
	\qquad \subfloat[]{\includegraphics[width=0.24\textwidth]{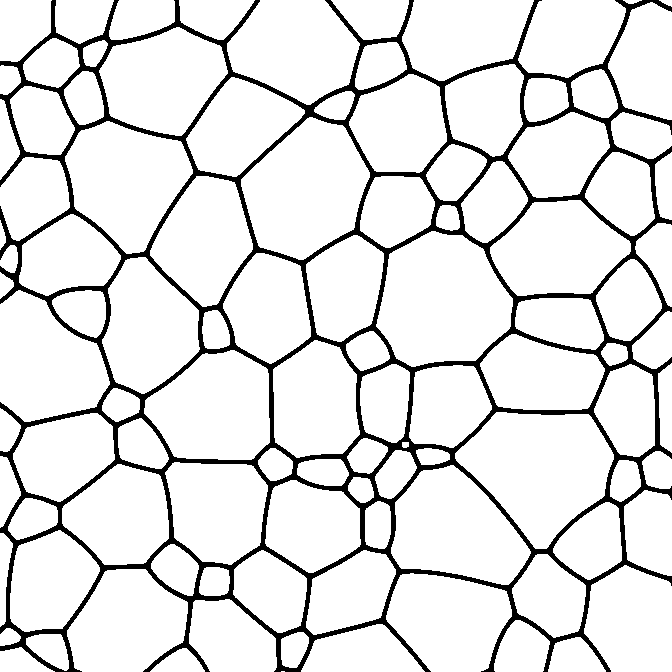}}
	\caption{Comparison between (a) experimental grains with (b) the general boundary structures of PFC and (c) Mullins. Note how PFC grains have meandering boundaries compared to the experimental ones, which have non-meandering boundaries that are straight in some cases and curved in others.}
	\label{fig:structures}
\end{figure}

Grain boundaries in polycrystalline materials are of paramount importance to various fields of science and engineering. As surrogates for difficult experimental investigations, several models have been developed to simulate grain growth and the evolution of the grain boundary network. These models include the mesoscale continuum model of grain growth stemming from Mullins' work \cite{MULLINS_Model} and the atomic to mesoscale Phase-Field-Crystal (PFC) models \cite{ELDER_Elasticity, EMMERICH_PFCReview}. A natural question is to what extent are such models capable of reproducing the morphology and dynamics of grain growth and the associated microstructural metrics such as the grain size distribution of real materials. Fig.~\ref{fig:structures} contrasts representative configurations of the grain boundary networks from experiments first presented in \cite{BARMAK_EarsTails} with PFC and Mullins-like simulations \cite{MULLINS_Model}. 
Although the grains produced by Mullins-like models appear visually more similar to experiments than the output of the PFC model, a definitive statement cannot be made in the absence of a quantitative comparison.

One key to answering this question is to exploit the universality of certain geometric statistics that are experimentally observed in polycrystalline materials, both in thin film and bulk form:  As the average grain size increases,  steady state distributions for many geometrical and topological properties are experimentally observed during the grain evolution \cite{FRADKOV_Experiments, MULLINS_Univ,  BARMAK_EarsTails, DONEGAN_Extreme}. These universal distributions give a precise measure of how to assess mathematical models via the extraction of the relevant metrics from simulations. In this vein, it was shown in \cite{BACKOFEN_GSD} that  the PFC  model provides a surprisingly accurate description of the grain size distribution, where the measure of ``grain size'' used was the reduced equivalent diameter. Previous work has also been done in computing certain statistical distributions for various Mullins-like models, see for example \cite{KINDERLEHRER_Variational, MASON_Experiments}.

In this article, we consider a  wide range of statistics, and present a detailed comparison of previously reported experimental results of thin film samples of aluminum \cite{BARMAK_EarsTails} with 2D simulations obtained from a basic PFC model and a well-known curvature driven grain boundary motion model \cite{ELSEY_NormalGrainGrowth}, here called the Mullins model. In all three cases, universal metric distributions are observed with respect to the dynamics and initial conditions. We demonstrate that for many though not all grain statistics, late time PFC gives better agreement with experiments than the Mullins model. However, the PFC model produces grain boundaries that meander far more than what is seen in experiments; its predictions related to grain shapes and topology therefore have lower accuracy. We present such a characterization of grain shape through a metric that we call the convex hull ratio. We remark that extracting data on grain structures from PFC runs is difficult because of the atomic nature of the model. This work benefits from the atom-based grain extraction procedure developed in \cite{MARTINE_AtomBased}, which forms a self-contained method whose accuracy has been tested against manual grain segmentations and the variational approach in \cite{HIRVONEN_Extract}. The current study underscores the critical importance of the use of multiple metrics for comparison of experiments and models for grain growth in polycrystalline materials, including amplitude equations based models \cite{GOLDENFELD_Amplitude}, more complicated PFC models, and the Potts' model \cite{PARAMDEEP_Potts}.

\section{Experiments and Simulations}
We first describe the origin of the three sources of data and the measurement of statistical grain measures or metrics. In all cases, the grains are represented on a finite domain so those lying on the boundary may be cut off. Such grains have been excluded from the analysis.

\subsection{Experimental data}
Experimental grain distributions have been obtained from imaging data in \cite{BARMAK_EarsTails}. Experimental details may be found there, but in short, aluminum thin films were prepared by sputter deposition then annealed at $400^\circ$C for a broad range of times (seconds to hours). The samples were then imaged via transmission electron microscopy using different viewing angles to characterize grain boundaries. These were traced manually, because with the exception of one sample \cite{BARMAK_EarsTails, CARPENTER_Errors}, automated grain identification had proven to be unsuccessful on account of the complex contrast of transmission electron micrographs. From these tracings, several geometric measures of the grain structure were obtained and presented in \cite{BARMAK_EarsTails}.

To extend the analysis of \cite{BARMAK_EarsTails} to some new grain metrics, all available original scanned hand tracings of the boundaries have been analyzed again. The scanned images were processed by identifying the grains as regions separated by boundaries a few pixels wide. The various metrics were then measured using simple image analysis routines. The analyzed subset consists of 86 images totaling roughly 3700 grains including as-deposited and annealed aluminum samples. The original set of more than 35000 original grains was reported in \cite{BARMAK_EarsTails} and included aluminum and copper thin films, but the manual hand tracings of the copper samples were not available for re-analysis and therefore could not be included in the current work.

We also note that the experimental grain size data of the more than 35000 grains included data from samples that had been annealed at a single time at a given temperature, as well as data from samples that had been examined as a function of time and showed that the structures were stagnant and did not evolve with time. Nevertheless, the grain size distributions of the partial and full datasets are similar, allowing the conclusion of a universal experimental grain size distribution in \cite{BARMAK_EarsTails}.

\subsection{PFC data}
The basic PFC model \cite{ELDER_Elasticity} can be written as the partial differential equation
\begin{equation*}
u_t = \nabla^2 \left( (\nabla^2 + 1)^2 u + u^3 - \beta u \right)
\end{equation*}
for a phase-field $u$ and a parameter $\beta$ that can be thought of as an inverse temperature. This is similar to the Swift-Hohenberg model \cite{SWIFT_Hohenberg} but the extra Laplacian on the right-hand side ensures that the average $\langle u \rangle = m$ remains fixed in time. In 2D, the phase diagram in the parameters $(m, \beta)$ can be divided into three regions: one where the constant state $u = m$ is stable, one where one-dimensional sinusoids are stable and finally a region where the hexagonal lattice is stable. In this regime, the model simulates the evolution of ``atoms'' represented by bumps in the phase field $u$. These bumps then arrange into a patchwork of hexagonal lattices with different orientations.

The PFC evolution was simulated with the scheme of \cite{ELSEY_Scheme} using periodic boundary conditions on a square shaped computational domain. Large-scale long-time grain statistics were extracted using the approach formulated in \cite{MARTINE_AtomBased}. As shown there and in \cite{HIRVONEN_Extract}, the procedure is reliable and agrees with manual grain segmentations and subsequent measurements. Since the timescale of PFC may be scaled by any arbitrary factor, the number of grains is a better indicator of progression than time. We extracted data when the domain of size $\approx 7429^2$ contained roughly 2200 and 135 grains, which we will label the Early and Late PFC distributions. As presented in \cite{MARTINE_AtomBased}, the PFC dynamics slows down significantly so Late PFC represents the ``universal'' distribution. We also note that to bypass any ambiguity concerning the grain perimeter, it is approximated by measuring the area of the disorded region between two grains and dividing by its thickness. More numerical details, including evidence of the robustness of PFC with respect to changing simulation parameters, may be found in the appendix.

In addition to grain area, perimeter, isoperimetric ratio (circularity) and the number of sides described in \cite{MARTINE_AtomBased}, we utilize an additional metric that we call the convex hull ratio:
\begin{equation*}
\frac{H - A}{A}
\end{equation*}
where $A$ and $H$ are respectively the area of the grain and of its convex hull, the smallest convex shape that fully encloses the grain. A value close to $0$ means the grain is almost convex while values far from 0 indicate that the grain is concave and ``meandering''. This metric is similar to the isoperimetric ratio in that it captures how far from a regular convex shape a grain is, but differs from it since it ignores grain elongation or equiaxedness. The convex hull ratio is straightforward to measure when grains are represented as regions. In the case of PFC images, the convex hull of a grain was computed by adding the area of the convex hull of its atoms, a computation similar to those described in \cite{MARTINE_AtomBased}. This agreed with projecting the grain onto a grid and using image analysis techniques.

\subsection{Mullins data}
We use a well-known, mesoscale, continuum model of curvature driven grain boundary motion that was developed by Mullins and others \cite{HARKER_Growth, BECK_Migration, MULLINS_Model}; see \cite{GOTTSTEIN_Textbook} for further references. In this context, grain boundaries are described as curves or surfaces undergoing geometric motion. There have been many algorithms proposed for the simulation of this and related models, including \cite{JANSSENS_Textbook} for some of the earlier approaches. We used threshold dynamics \cite{MERRIMAN_MultipleJunctions}, a recent, level set type algorithm that is particularly efficient due to its unconditional stability, allowing simulations of up to hundreds of thousands of grains in  both two and three dimensions \cite{ESEDOGLU_Multiphase, ELSEY_NormalGrainGrowth}. Versions of this algorithm exist for very general (including normal dependent) surface tensions and mobilities. For the simulations reported in this paper, referred to as the Mullins model, we took all surface tensions and mobilities to be isotropic and equal, resulting in symmetric dihedral angles of $120^\circ$ at triple junctions, and carried out the simulations in 2D in order to compare with PFC and the thin film experiments. Periodic boundary conditions were used on a square shaped computational domain. The random Voronoi initial condition contained in excess of 100000 grains, which were then allowed to coarsen down to 11910 final grains. The final grain structure was analyzed using image analysis techniques. We note that the extracted metrics are stable after an initial transient.

\section{Comparison}
We now compare the statistical grain metrics. The grain distributions contained 3678, 81343, 4937, and 11910 data points for the experimental, Early PFC, Late PFC and Mullins distributions respectively. The histograms below are constructed by choosing the bin count with the Freedman-Diaconis rule \cite{FREEDMAN_Histogram} and error bars represent standard errors. For histograms, this error is approximately $\sqrt{n_i}$ where $n_i$ is the count of each bin, assuming that datapoints are chosen randomly from their underlying distribution. This is an approximation as grains are necessarily correlated, but this effect is small if the domain is larger than the scale of such correlations, which should be on the order of a few grain widths.

Experimental errors are much more difficult to assess precisely. For example, in all three cases, there is an error associated to measuring grain metrics once the boundaries have been identified, but it is small since the domain's discretization is small compared to the size of grains. In experimental images (see \cite{CARPENTER_Errors} for a discussion) and PFC, there is an additional error in properly characterizing these boundaries, meaning that some grains may not be properly identified. Such errors are quite small since ambiguities are present near boundaries which make up a relatively small fraction of the domains. Finally, recognizing PFC boundaries is complicated by their atomic nature, but it was evaluated in \cite{MARTINE_AtomBased} that discrepancies in measurements between numerical and manual segmentations amount to a few percent. Overall, the relatively small number of datapoints mean that the statistical errors overshadow experimental errors, so the error bars presented below can help guide the eye as to the closeness of the compared data.

\subsection{Reduced area and reduced equivalent circle diameter}
\begin{figure}[h]
	\centering
	\subfloat[]{\includegraphics[width=0.5\textwidth]{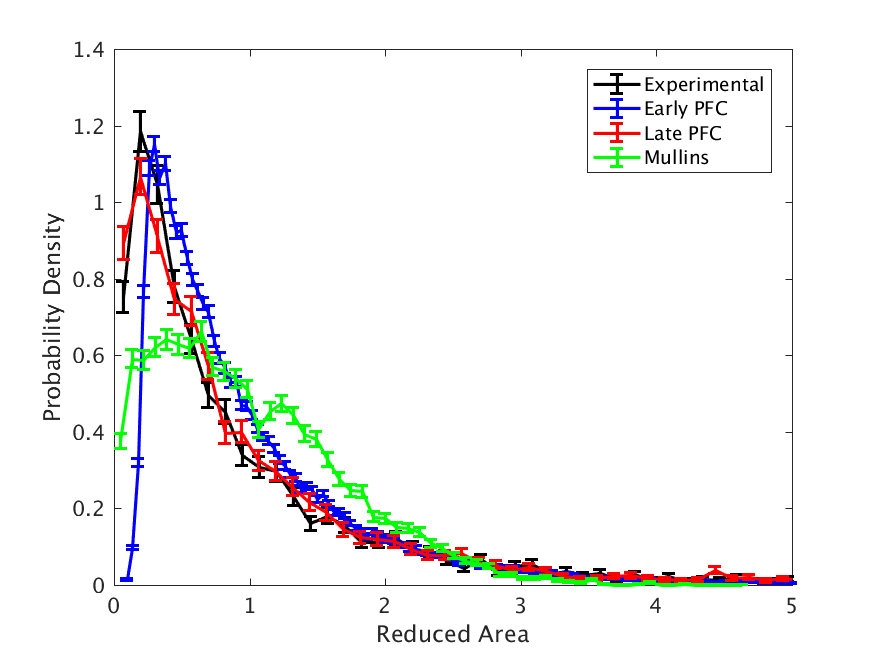}}
	\subfloat[]{\includegraphics[width=0.5\textwidth]{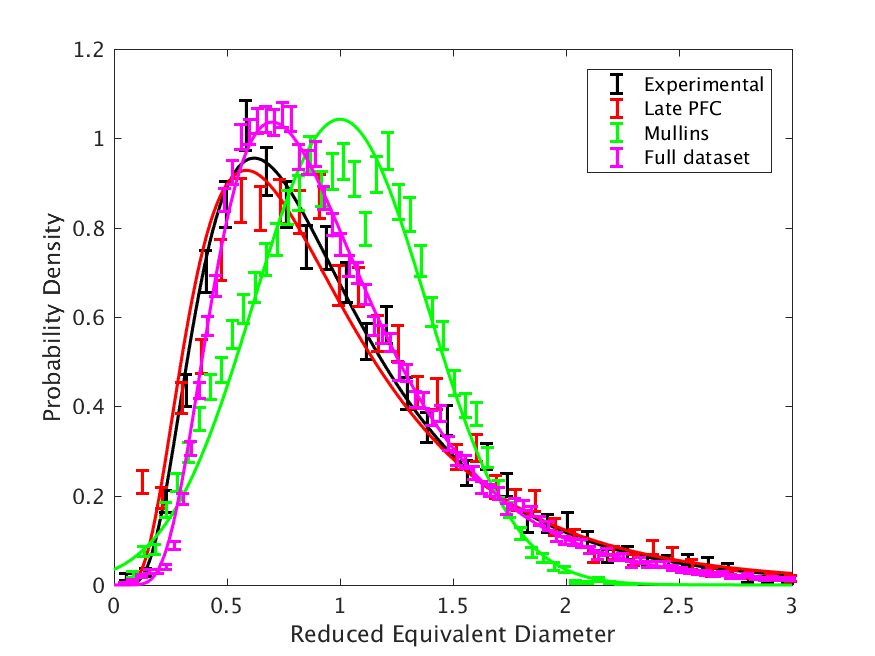}}\\
	\caption{(a) Grain area distribution, normalized with respect to the average. (b) Equivalent diameter distribution, normalized with respect to the average. The smooth curves correspond to the fits in table \ref{tab:lognormal}. The Early PFC distribution has been omitted for clarity while the full experimental distribution has been included for comparison.}
	\label{fig:statistics1}
\end{figure}

Given an area $A$ for a grain, the reduced area distribution is given by the set $A/\langle A \rangle$ where $\langle A \rangle$ denotes the average area. A derived metric, the equivalent circle diameter $D$, corresponds to the diameter of the circle with area $A$. In Fig.~\ref{fig:statistics1}(a) and (b), we see that both the area and the diameter distributions clearly show that the Mullins model behaves quite differently from experimental grains and PFC. The main feature compared to random tesselations, as noted in \cite{BARMAK_EarsTails, BACKOFEN_GSD}, is the presence of a population of small grains (the region in the distribution termed ``ear'') and a population of very large grains (the region termed the ``tail''). The experimental distributions lies close to the Late PFC distributions. Another point is that the reduced equivalent diameter distributions are lognormal in the PFC and experimental cases, but almost normal for Mullins, highlighting a drastically different behavior. The reduced equivalent diameter distributions were fit to the appropriate distributions and the fit parameters are reproduced in Table \ref{tab:lognormal}. The conclusion here is similar: PFC moves towards the experimental data. We emphasize that PFC has moved slightly past the full experimental dataset.

\begin{table}[h]
\caption{\label{tab:lognormal}Fit parameters for the different datasets. Experiments and PFC fits are lognormal so $(\mu, \sigma)$ refer to the average and standard deviation of the logarithm. The Mullins fit is normal so the parameters are the usual average and standard deviation and should not be compared with lognormal parameters. The fit reduced average is computed from the fit parameters and can be compared to the data reduced average of $1$.}
\center
\begin{tabular}{cccc}
									&	$\mu$	&	$\sigma$	&	Fit reduced average\\ \hline
Experimental (3700 grains subset)	&	-0.15	&	0.57		&	1.012\\
Experimental (35000 grains set)		&	-0.12	&	0.49		&	1.001\\
Early PFC							&	-0.07	&	0.38		&	0.998\\
Late PFC							&	-0.16	&	0.61		&	1.022\\
Mullins (normal fit)				&	1.00	&	0.38		&	1.000\\
\end{tabular}
\end{table}

\subsection{Reduced perimeter and isoperimetric Ratio}
\begin{figure}[h]
	\centering
	\subfloat[]{\includegraphics[width=0.5\textwidth]{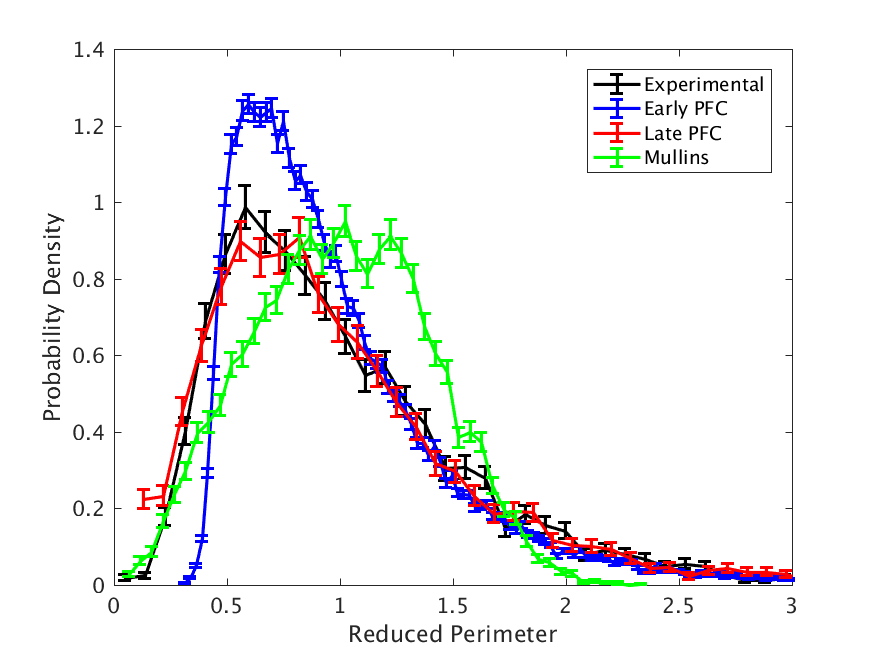}}
	\subfloat[]{\includegraphics[width=0.5\textwidth]{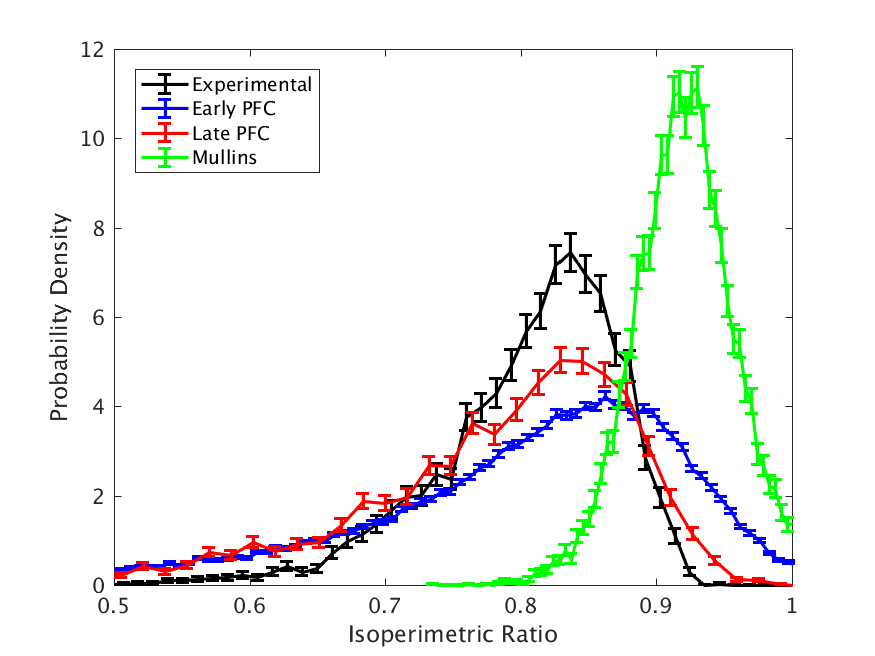}}\\
	\caption{(a) Grain perimeter distribution, normalized with respect to the average. (b) Grain isoperimetric ratio. The isoperimetric ratio is defined as $4\pi A/P^2$ where $A, P$ are the area and perimeter of a grain.}
	\label{fig:statistics2}
\end{figure}

As with area, Fig.~\ref{fig:statistics2}(a) shows excellent agreement between the experimental and PFC results for the reduced perimeter distribution. A quantity derived from area and perimeter is the isoperimetric ratio, also called the circularity. This quantity measures how close a grain is to a circle and thus how round or compact it is. This is similar to other equiaxial metrics such as the aspect ratio. Fig.~\ref{fig:statistics2}(b) shows that both experimental distributions and PFC are broad while Mullins presents a sharply peaked distribution. 

\subsection{Convex Hull Ratio}
\begin{figure}[h]
	\centering
	\includegraphics[width=0.5\textwidth]{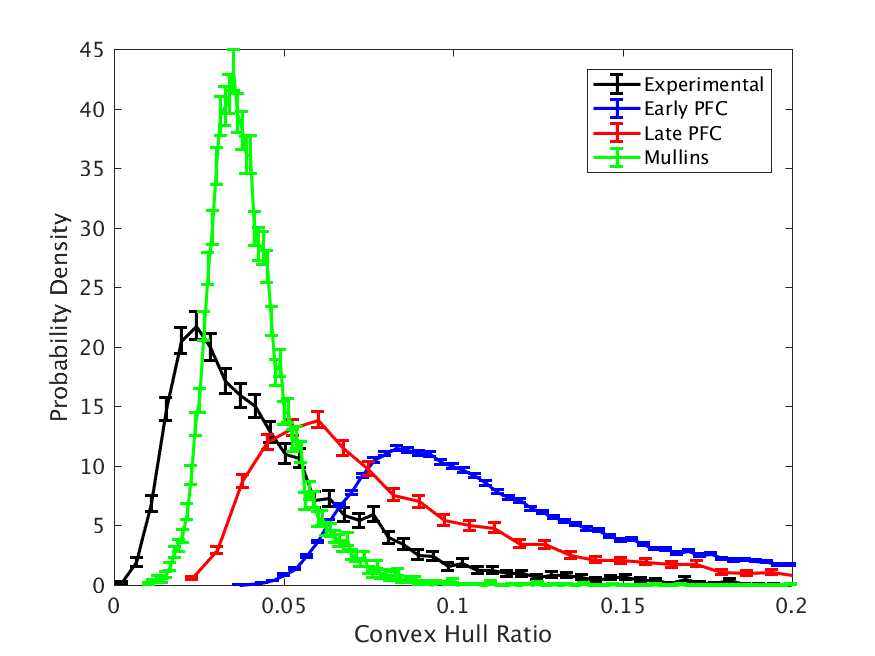}
	\caption{Grain convex hull ratio, defined as $(H-A)/A$ where $A$ and $H$ are the area of the grain and of its convex hull, respectively.}
	\label{fig:statistics4}
\end{figure}

Fig.~\ref{fig:statistics4} shows the convex hull ratio, equal to 0 for convex grains and positive for concave and meandering grain boundaries. The comparison shows that the experimental distribution is quite far from both PFC and Mullins yet shares a similar shape to PFC. Mullins on the other hand is very sharply peaked and away from 0, so most of its grains have slightly curved boundaries. The main difference here is that PFC produces grains with meandering boundaries while experimental and Mullins grains presents roughly polygonal grains but with slightly curved boundaries. However, the distribution of curvatures is much broader in experiments than in the Mullins model. For PFC, this metric has not stabilized to the same extent as the others, however, its evolution also slows down in a similar manner so that it is highly unlikely that it would eventually reach to the experimental distribution.

\subsection{Metrics based on the number of sides distribution}
\begin{figure}[H]
	\centering
	\subfloat[]{\includegraphics[width=0.5\textwidth]{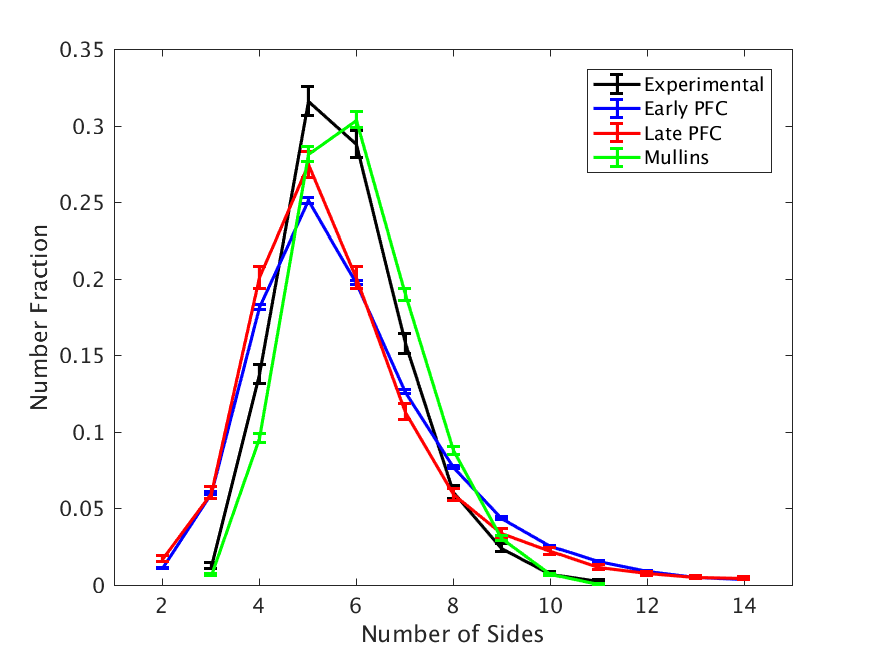}}
	\subfloat[]{\includegraphics[width=0.5\textwidth]{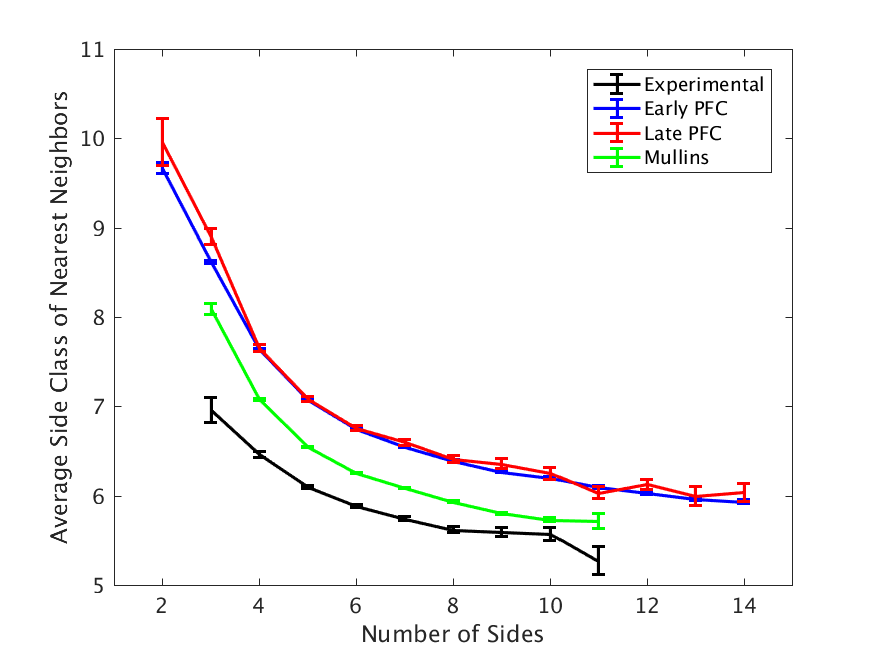}}
	\\
	\subfloat[]{\includegraphics[width=0.5\textwidth]{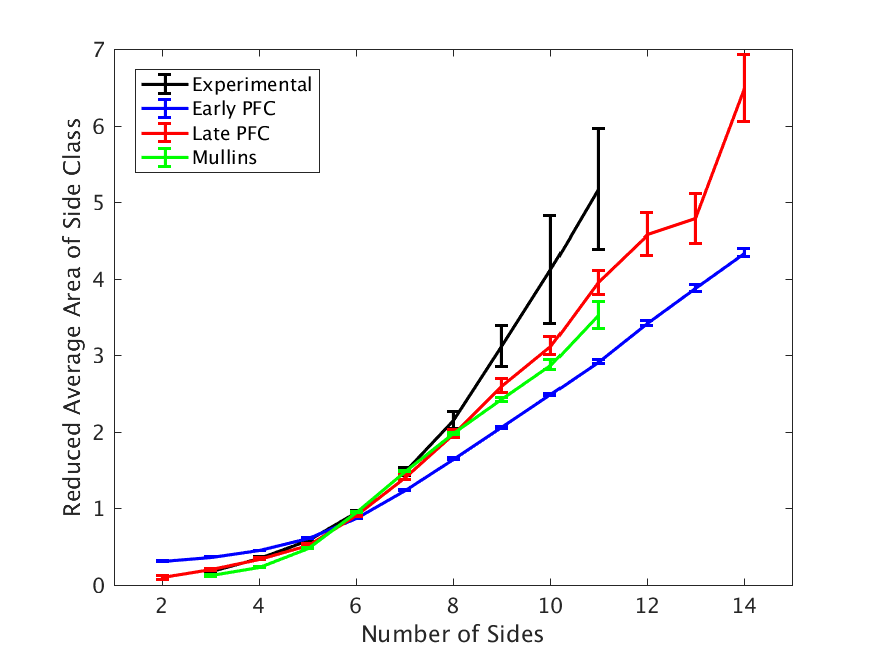}}
	\subfloat[]{\includegraphics[width=0.5\textwidth]{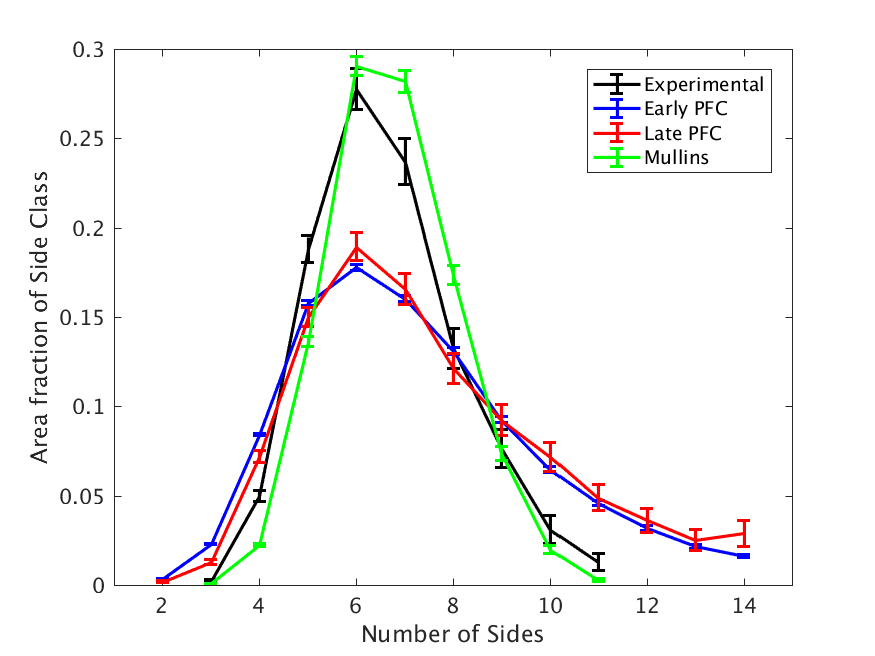}}
	\caption{(a) Number of sides distribution computed by counting the number of neighbors of a grain. (b) Average number of sides of nearest neighbors binned by the number of sides. This quantity is computed per grain by taking its neighbors and averaging their own number of sides. (c) Reduced average area corresponding to each side class. Fix a side class and compute the average reduced area of grains in this side class. (d) Area fraction of grains in a given side class. This quantity is computed by multiplying the number of sides distribution with the previous reduced average area. Connecting lines are drawn to guide the eye and standard errors were computed and are plotted in (b), (c), (d).}
	\label{fig:statistics3}
\end{figure}

The number of sides of a grain, called its side class, is defined as its number of neighbors. In contrast to the other results, Fig.~\ref{fig:statistics3}(a) shows that the number of sides distribution of the experimental distribution is closer to Mullins than to PFC, which exhibits a number of grains with very few or many neighbors. Nevertheless, it appears that both the experimental distributions and PFC peak at 5 neighbors rather than 6.

Several other metrics combining geometry and topology can be computed. Fig.~\ref{fig:statistics3}(b) shows the average number of sides of the nearest neighbors for a given side class, which shows that grains with few sides have neighbors with many neighbors and vice-versa. The average reduced area for a given side class is presented in Fig.~\ref{fig:statistics3}(c), showing that grains with many neighbors occupy a larger fraction of the domain than those with few neighbors. In other words, large grains have many neighbors.

These two metrics have been compared respectively to the Aboav-Weaire and Lewis laws in detail in \cite{BARMAK_EarsTails} and the reader is directed to \cite{CHIU_AboavWeaireLewis} for physical details. In short, the Aboav-Weaire law predicts that the curves in (b) would follow $5 + 8/N$ where $N$ is the side class. This law, observed across a variety of tesselated systems, can be shown to hold rigorously if the grains correspond to a random convex tesselation of the domain. The Mullins model matches this law quite well while experiments and PFC match the law with different constants but keep the same trend. This makes it clear that PFC grains have on average more grains than would be expected from random tesselations, most likely due to the non-convexity of PFC grains. On the other hand, the Lewis law predicts a linear relationship in (c) which does not appear to be followed by any of the comparison models.

Lastly, the area fraction of grains with a given side class is shown in Fig.~\ref{fig:statistics3}(d). This metric is quite similar to the number of sides distribution and shows more clearly than (a) that experimental data is closer to Mullins while PFC is stationary away from the other two. Overall, these three different metrics of grain structure show that there are important discrepancies in comparing PFC to experimental data from a topological point of view.

\section{Discussion and conclusion}
The comparisons above show that the PFC model is more successful than Mullins simulations at predicting the purely geometric metrics of grain. In particular, the grain size distribution of late time PFC agrees well with that obtained from experimental data of aluminum films, as highlighted in \cite{BACKOFEN_GSD}. However, as suggested by visual comparisons (see Fig.~\ref{fig:structures}), those metrics related to grain shape and topology are much less accurately captured by PFC. Indeed, our comparison via the convex hull ratio suggests that a significant deficiency of the PFC model is its propensity to form meandering grain boundaries, which does not appear to resolve in time. On the other hand, while the Mullins model accurately resolves the number of sides distribution, it fails to capture most other measures.

These findings reiterate many important questions as to universality and its inherent presence in the models. First off, is this universality preserved across all polycrystalline materials with respect to dimensionality or lattice structure (FCC, BCC, Cubic, etc)? Secondly, how is PFC able to predict the distributions for metrics related to area and perimeter? The question is difficult to fully answer, but perhaps a naive answer is as follows.

PFC is the simplest conservative PDE that may give rise to polycrystalline patterns formed by individual atoms as it encodes only the core of a two-body correlation function: neighboring atoms trying to be a given fixed distance from each other \cite{EMMERICH_PFCReview}. In fact, this suggests that the main character of the basic PFC model is close packing, producing hexagonal lattices in 2D. As such, the mechanism of grain evolution must rely mostly on neighbor to neighbor interactions. Perhaps the apparent agreement of PFC with experiments indicates that the deciding mechanism for grain size in real materials is simply close packing. This naive interpretation may help explain why our 2D Hexagonal PFC results match the 3D FCC aluminum measurements of thin-films. On the other hand, the main limitation of PFC is that it forms meandering boundaries with highly concave grains, meaning that PFC lacks a driving force to promote the grain convexity that is clearly observed in experimental grains. In contrast, the Mullins model is driven by minimization of the curvature of grain boundaries, so it is not surprising that it is more successful to characterize grain shape. Conversely, it fails to recover the underlying atomic nature of grains since it cannot represent any other defect.

It would be interesting to see if the naive intuition that characterizes PFC as close packing could be extended to other models capable of simulating different lattice types. Would introducing corrections such as orientation dependent surface tensions in the basic Mullins model allow it to better reproduce experimental data? Lastly, an important mathematical question is to what extend one can analytically capture such universal behavior directly from the models?

To conclude, we have compared important geometric metrics of the grain distributions of experimental structures with the PFC and Mullins models and found that they better capture geometric and shape metrics, respectively.

\section*{Acknowledgments}
GMLaB was supported by an FRQNT (Fonds de recherche du Qu\'ebec - Nature et technologies) Doctoral Student Scholarship. RC was supported by an NSERC (Natural Sciences and Engineering Research Council Canada) Discovery Grant. KB gratefully acknowledges financial support of the SRC, Task 1292.008, 2121.001, 2323.001 and of the MRSEC programs of the US NSF under DMR-0520425. SE was supported by the US NSF under DMS-1719727. We thank the anonymous referees for comments that improved this manuscript.

\section*{Declaration of interest}
The authors declare no conflict of interest.

\section*{Appendix: Details on PFC simulations and statistics}
The PFC results were obtained by simulating the PFC evolution with the scheme of \cite{ELSEY_Scheme} using the same parameters $(m, \beta) = (0.07, 0.025)$, $C = 2\beta = 0.05$ and $\tau = 1000$ for a domain of size $L \approx 7429$ with $8192$ grid points. Since the interatomic distance between two atoms equals $2\pi/\sqrt{3}$, this domain size supports roughly 1024 atoms, each resolved by about 8 grid points (8 pixels per atom), in each direction. The evolution was run for 40000 time steps starting with 36 random initial phase fields. Data was first captured when the domain contained roughly 2200 grains (Early PFC) and the simulations ended with an average of 135 grains (Late PFC). The grain extraction procedure was carried out using the parameters found optimal in \cite{MARTINE_AtomBased}, $h = -0.035$, $\gamma = 0.001$, $\theta = 2.5^\circ$ and $\alpha = 40$.

\begin{figure}[H]
	\centering
	\subfloat[]{\includegraphics[width=0.32\textwidth]{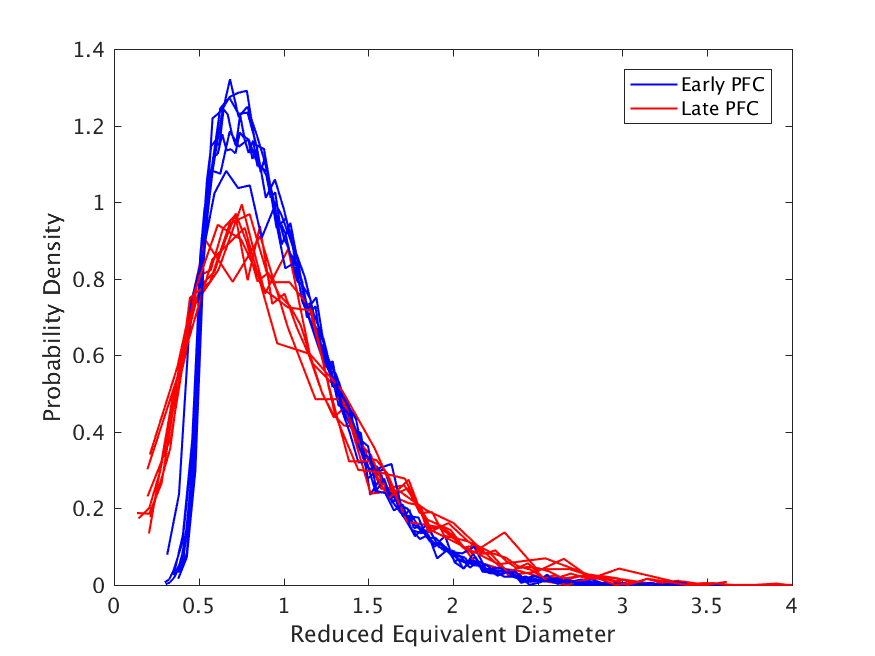}}
	\subfloat[]{\includegraphics[width=0.32\textwidth]{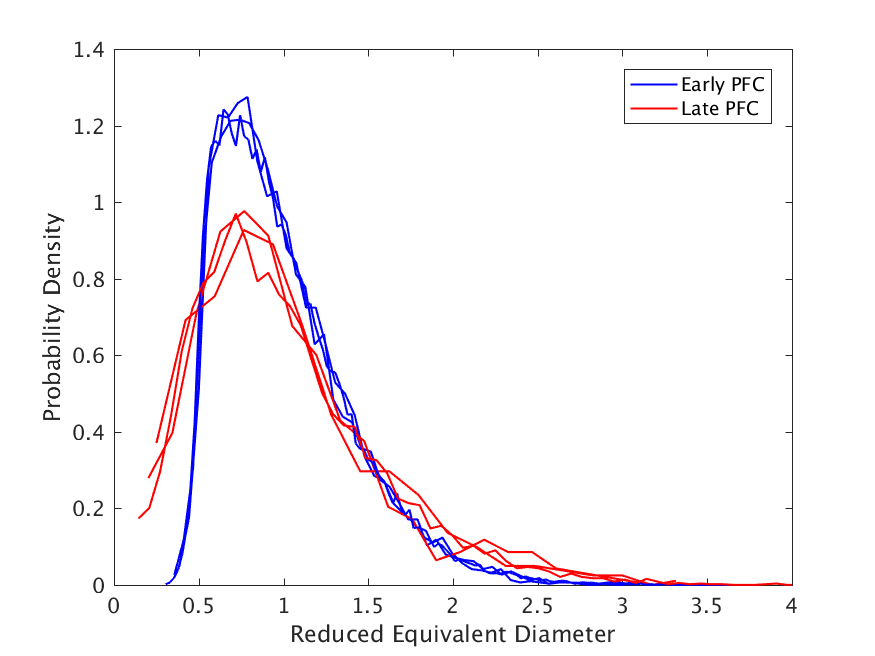}}
	\subfloat[]{\includegraphics[width=0.32\textwidth]{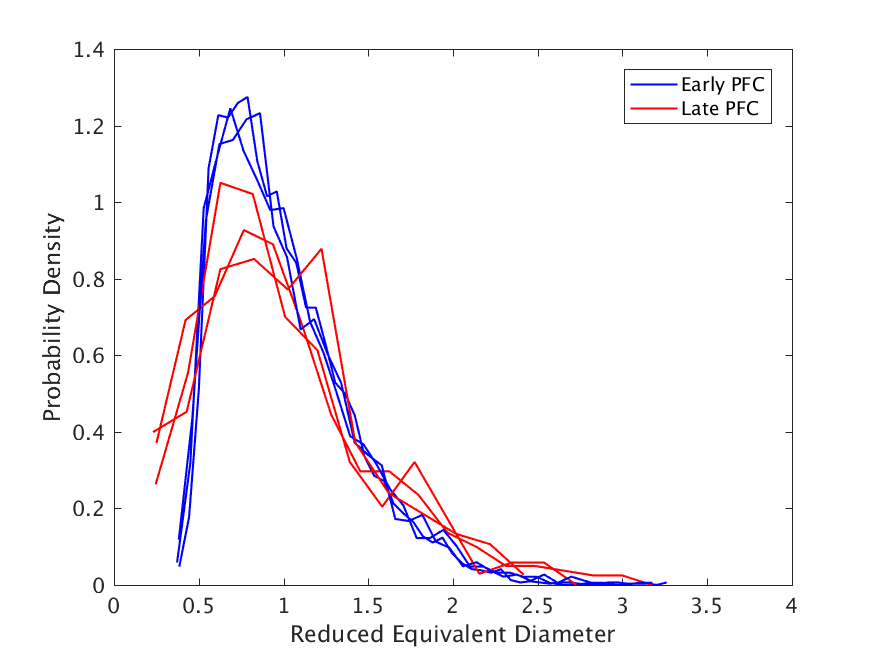}}
	\caption{Comparison between different PFC input parameters, varying the PFC parameters (a), the domain size (b) and the grid resolution (c). Variability can be gauged by the spread in the histograms; the PFC evolution then gives rise to robust statistics that are not very sensitive to simulation details.}
	\label{fig:paramcomparison}
\end{figure}

We briefly compare the statistics obtained by varying the PFC parameters, the domain size and the grid resolution, showing that the PFC model gives rise to robust statistics over a variety of input parameters. The comparison is shown for the reduced equivalent diameter distribution but other metrics behave similarly. In Fig.~\ref{fig:paramcomparison}(a), we varied the PFC parameters $(m, \beta)$ away from $(0.07, 0.025)$ to include $(0.05, 0.016)$, $(0.06, 0.025)$, $(0.07, 0.016)$, $(0.07, 0.034)$, $(0.08, 0.025)$ and $(0.09, 0.034)$. These lie within the hexagonal regime close to the order disorder transition curve \cite{ELDER_Elasticity}. The parameters $C$ and $h$ were varied to accommodate the resulting atoms but the numerical domain was kept fixed. In Fig.~\ref{fig:paramcomparison}(b), we varied the domain size to fit $256^2, 512^2$ and $1024^2$ atoms, keeping the grid resolution fixed at about $8^2$ pixels per atom. In Fig.~\ref{fig:paramcomparison}(c), we varied the grid resolution from $8^2, 16^2$ to $32^2$ pixels per atoms while keeping the domain size fixed to fit $256^2$ atoms.

\bibliographystyle{elsarticle-num}
\bibliography{References}

\begin{thebibliography}{10}
\expandafter\ifx\csname url\endcsname\relax
  \def\url#1{\texttt{#1}}\fi
\expandafter\ifx\csname urlprefix\endcsname\relax\def\urlprefix{URL }\fi
\expandafter\ifx\csname href\endcsname\relax
  \def\href#1#2{#2} \def\path#1{#1}\fi

\bibitem{MULLINS_Model}
W.~W. Mullins, Two dimensional motion of idealized grain boundaries, Journal of
  Applied Physics 27~(8) (1956) 900--904.
\newblock \href {https://doi.org/10.1063/1.1722511}
  {\path{doi:10.1063/1.1722511}}.

\bibitem{ELDER_Elasticity}
K.~R. Elder, M.~Katakowski, M.~Haataja, M.~Grant, Modeling elasticity in
  crystal growth, Physical Review Letters 88 (2002) 245701.
\newblock \href
  {https://doi.org/http://dx.doi.org/10.1103/PhysRevLett.88.245701}
  {\path{doi:http://dx.doi.org/10.1103/PhysRevLett.88.245701}}.

\bibitem{EMMERICH_PFCReview}
H.~{Emmerich}, H.~{L{\"o}wen}, R.~{Wittkowski}, T.~{Gruhn}, G.~I. {T{\'o}th},
  G.~{Tegze}, L.~{Gr{\'a}n{\'a}sy}, Phase-field-crystal models for condensed
  matter dynamics on atomic length and diffusive time scales: an overview,
  Advances in Physics 61~(6) (2012) 665--743.
\newblock \href
  {https://doi.org/http://dx.doi.org/10.1080/00018732.2012.737555}
  {\path{doi:http://dx.doi.org/10.1080/00018732.2012.737555}}.

\bibitem{BARMAK_EarsTails}
K.~Barmak, E.~Eggeling, D.~Kinderlehrer, R.~Sharp, S.~Ta’asan, A.~Rollett,
  K.~Coffey, Grain growth and the puzzle of its stagnation in thin films: The
  curious tale of a tail and an ear, Progress in Materials Science 58~(7)
  (2013) 987 -- 1055.
\newblock \href
  {https://doi.org/http://dx.doi.org/10.1016/j.pmatsci.2013.03.004}
  {\path{doi:http://dx.doi.org/10.1016/j.pmatsci.2013.03.004}}.

\bibitem{FRADKOV_Experiments}
V.~E. Fradkov, A.~S. Kravchenko, L.~S. Shvindlerman, Experimental investigation
  of normal grain growth in terms of area and topological class, Scripta
  metallurgica 19~(11) (1985) 1291--1296.

\bibitem{MULLINS_Univ}
W.~W. Mullins, The statistical self-similarity hypothesis in grain growth and
  particle coarsening, Journal of Applied Physics 59 (1986) 1341.

\bibitem{DONEGAN_Extreme}
S.~P. Donegan, J.~C. Tucker, A.~D. Rollett, K.~Barmak, M.~Groeber, Extreme
  value analysis of tail departure from log-normality in experimental and
  simulated grain size distributions, Acta Materialia 61~(15) (2013)
  5595--5604.
\newblock \href
  {https://doi.org/http://dx.doi.org/10.1016/j.actamat.2013.06.001}
  {\path{doi:http://dx.doi.org/10.1016/j.actamat.2013.06.001}}.

\bibitem{BACKOFEN_GSD}
R.~Backofen, K.~Barmak, K.~Elder, A.~Voigt, Capturing the complex physics
  behind universal grain size distributions in thin metallic films, Acta
  Materialia 64 (2014) 72 -- 77.
\newblock \href
  {https://doi.org/http://dx.doi.org/10.1016/j.actamat.2013.11.034}
  {\path{doi:http://dx.doi.org/10.1016/j.actamat.2013.11.034}}.

\bibitem{KINDERLEHRER_Variational}
D.~Kinderlehrer, I.~Livshits, S.~Ta’Asan, A variational approach to modeling
  and simulation of grain growth, SIAM Journal on Scientific Computing 28~(5)
  (2006) 1694--1715.

\bibitem{MASON_Experiments}
J.~K. Mason, E.~A. Lazar, R.~D. MacPherson, D.~J. Srolovitz, Geometric and
  topological properties of the canonical grain-growth microstructure, Phys.
  Rev. E 92 (2015) 063308.
\newblock \href {https://doi.org/10.1103/PhysRevE.92.063308}
  {\path{doi:10.1103/PhysRevE.92.063308}}.

\bibitem{ELSEY_NormalGrainGrowth}
M.~Elsey, S.~Esedoḡlu, P.~Smereka, Large-scale simulation of normal grain
  growth via diffusion-generated motion, Proceedings of the Royal Society of
  London A: Mathematical, Physical and Engineering Sciences\href
  {https://doi.org/10.1098/rspa.2010.0194} {\path{doi:10.1098/rspa.2010.0194}}.

\bibitem{MARTINE_AtomBased}
G.~Martine La~Boissoniere, R.~Choksi, Atom based grain extraction and
  measurement of geometric properties, Modelling and Simulation in Materials
  Science and Engineering 26~(3) (2018) 035001.
\newblock \href {https://doi.org/http://dx.doi.org/10.1088/1361-651X/aaa635}
  {\path{doi:http://dx.doi.org/10.1088/1361-651X/aaa635}}.

\bibitem{HIRVONEN_Extract}
P.~Hirvonen, G.~Martine La~Boissoni\`ere, Z.~Fan, C.~Achim, N.~Provatas, K.~R.
  Elder, T.~Ala-Nissila, Grain extraction and microstructural analysis method
  for two-dimensional poly and quasicrystalline solids, Phys. Rev. Materials 2
  (2018) 103603.
\newblock \href {https://doi.org/10.1103/PhysRevMaterials.2.103603}
  {\path{doi:10.1103/PhysRevMaterials.2.103603}}.

\bibitem{GOLDENFELD_Amplitude}
N.~Goldenfeld, B.~P. Athreya, J.~A. Dantzig, Renormalization group approach to
  multiscale simulation of polycrystalline materials using the phase field
  crystal model, Physical Review E 72~(2) (2005) 020601.
\newblock \href {https://doi.org/http://dx.doi.org/10.1103/PhysRevE.72.020601}
  {\path{doi:http://dx.doi.org/10.1103/PhysRevE.72.020601}}.

\bibitem{PARAMDEEP_Potts}
P.~S. Sahni, G.~S. Grest, M.~P. Anderson, D.~J. Srolovitz, Kinetics of the
  $q$-state potts model in two dimensions, Phys. Rev. Lett. 50 (1983) 263--266.
\newblock \href {https://doi.org/10.1103/PhysRevLett.50.263}
  {\path{doi:10.1103/PhysRevLett.50.263}}.

\bibitem{CARPENTER_Errors}
D.~T. Carpenter, J.~M. Rickman, K.~Barmak, A methodology for automated
  quantitative microstructural analysis of transmission electron micrographs,
  Journal of applied physics 84~(11) (1998) 5843--5854.

\bibitem{SWIFT_Hohenberg}
J.~Swift, P.~C. Hohenberg, Hydrodynamic fluctuations at the convective
  instability, Phys. Rev. A 15 (1977) 319--328.
\newblock \href {https://doi.org/10.1103/PhysRevA.15.319}
  {\path{doi:10.1103/PhysRevA.15.319}}.

\bibitem{ELSEY_Scheme}
M.~Elsey, B.~Wirth, A simple and efficient scheme for phase field crystal
  simulation, ESAIM: Mathematical Modelling and Numerical Analysis 47~(5)
  (2013) 1413–1432.
\newblock \href {https://doi.org/http://dx.doi.org/10.1051/m2an/2013074}
  {\path{doi:http://dx.doi.org/10.1051/m2an/2013074}}.

\bibitem{HARKER_Growth}
D.~Harker, E.~R. Parker, Grain shape and grain growth, Transactions of the
  American Society for Metals 34 (1945) 156--201.

\bibitem{BECK_Migration}
P.~Beck, Interface migration in recrystallization, Metal interfaces (1952)
  208--247.

\bibitem{GOTTSTEIN_Textbook}
G.~Gottstein, L.~S. Shvindlerman, Grain boundary migration in metals:
  thermodynamics, kinetics, applications, CRC press, 2009.

\bibitem{JANSSENS_Textbook}
K.~G.~F. Janssens, D.~Raabe, E.~Kozeschnik, M.~A. Miodownik, B.~Nestler,
  Computational materials engineering: An introduction to microstructure
  evolution. 1st edition, Elsevier. London, UK, 2007.

\bibitem{MERRIMAN_MultipleJunctions}
B.~Merriman, J.~K. Bence, S.~J. Osher, Motion of multiple junctions: A level
  set approach, Journal of Computational Physics 112~(2) (1994) 334 -- 363.
\newblock \href {https://doi.org/https://doi.org/10.1006/jcph.1994.1105}
  {\path{doi:https://doi.org/10.1006/jcph.1994.1105}}.

\bibitem{ESEDOGLU_Multiphase}
S.~Esedoḡlu, S.~Ruuth, R.~Tsai, Diffusion generated motion using signed
  distance functions, Journal of Computational Physics 229 (2010) 1017--1042.
\newblock \href {https://doi.org/10.1016/j.jcp.2009.10.002}
  {\path{doi:10.1016/j.jcp.2009.10.002}}.

\bibitem{FREEDMAN_Histogram}
D.~Freedman, P.~Diaconis, On the histogram as a density estimator: L 2 theory,
  Probability theory and related fields 57~(4) (1981) 453--476.

\bibitem{CHIU_AboavWeaireLewis}
S.~N. Chiu, Aboav-weaire's and lewis' laws - a review, Materials
  characterization 34~(2) (1995) 149--165.

\end{thebibliography}

\end{document}